\documentclass{article}

\usepackage{graphicx}

\begin{document}

\title{\LARGE{Utilization of electromagnetic detector for selection
and detection of high-frequency relic gravitational waves}\footnote{
Supported by the National Basic Research Programmer of China under
Grant No 2003CB716300, the Natural Science Foundation of Chongqing
under Grant No 8562, and the National Natural Science Foundation of
China under Grant No 10575140.}}

\author{\small Fangyu Li \footnote{ Email:
fangyuli@cqu.edu.cn},  Zhenya Chen, Ying Yi  \\
\small Department of Physics, Chongqing University, Chongqing
400044}

\maketitle

\begin{abstract}

\textbf{It is shown that coupling system between fractal membranes
and a Gaussian beam passing through a static magnetic field, has
strong selection capability for the stochastic relic gravitational
wave (GW) background. The relic GW components propagating along
the positive direction of the symmetrical axis of the Gaussian
beam, might generate an optimal electromagnetic perturbation,
while the perturbation produced by the relic GW components
propagating along the negative and perpendicular directions to the
symmetrical axis, will be much less than the former, and the
influence of the random fluctuation of the relic GWs to such
effect, can be neglected. The high-frequency relic GWs satisfying
the parameters requirement (\textit{h}{$\sim $10}$^{ - 31}$ or
larger), frequency resonance and ``direction coupling'', in
principle, would be selectable and measurable in seconds.}
\\


\end{abstract}

In resent years, whether quintessential inflationary models (QIM)
[1-3] or some string cosmology scenarios [4-7], they all predicted a
high energy density region of relic gravitons in the microwave band
($\sim 10^9 - 10^{11}$Hz) (although there are some critiques to the
scenarios, whether they have uncovered a fatal flaw in the scenarios
remains to be determined), the corresponding dimensionless amplitude
of the relic gravitational waves (GWs) in the region, may reach up
to roughly $ \sim 10^{ - 30} - 10^{ - 32}$ [1,2,7]. Since such
frequencies are just the best electromagnetic (EM) detecting band
for the high-frequency gravitational waves (HFGWs), it caused
extensive interest and reviews [1,7-11].

In our some previous works [10-12], we considered the resonant
response of a special EM system to the HFGWs. The EM system consists
of new-type fractal membranes [13,14] and a Gaussian beam passing
through a static magnetic field, and it is found that if the HFGW
propagates along the positive direction of the symmetrical axis (the
$z$-axis) of the Gaussian beam, and satisfies resonant condition
(the frequency \textit{$\nu $}$_{e}$ of the Gaussian beam is tuned
to the frequency \textit{$\nu $}$_{g}$ of the GW), it will generate
an optimum resonant response, and the EM perturbation has a good
space accumulation effect in the propagating direction of the GW.
However, since the random property of the relic GWs, the propagating
directions of the relic GWs are also stochastic, and because of
stochastic fluctuation of the amplitudes of the relic GWs, detection
of the relic GWs will be more difficult than that of the
monochromatic plan GWs. In this case, can we select and measure them
by the EM detectors? In particular, if both the high-frequency relic
GWs have the same amplitude and frequency, but propagate along the
opposite directions (standing wave), may their effect be
counteracted for each other? As we shall show that in our EM system
the EM perturbations produced by the relic GWs propagating along the
positive and negative directions of the $z$-axis, \textit{will be
non-symmetric}. Moreover, the physical effect generated by the relic
GWs propagating along other directions, would be also \textit{quite
different}, even if they satisfy the resonant condition ($\omega _e
= \omega _g )$, and only the relic GW component propagating along
the positive direction of the symmetrical axis of the Gaussian beam,
can generate an optimal resonant response. Thus our EM system would
be very sensitive to the propagating directions and frequency of the
relic GWs.

It is well known that each polarization component $h(\eta ,\vec {x})$ of the
relic GW can be written as [1, 2, 15]
\\

\begin{equation}
\label{eq1}
h(\eta ,\vec {x}) = \frac{\mu (\eta )}{a}\exp (i\vec {k} \cdot \vec {x}),
\end{equation}
\\

\noindent the time dependence of $h$ is determined by the $\mu (\eta
)$ satisfying the equation
\\

\begin{equation}
\label{eq2}
{\mu }'' + (k^2 - {a}'' / a)\mu = 0,
\end{equation}
\\

\noindent where ${a}'' = \frac{\partial ^2a}{\partial \eta ^2}$, $a
= a(\eta )$ is the cosmology scale factor, $\eta $ is the conformal
time. For the relic GWs of the very high-frequency in the GHz band
(i.e., the gravitons of large momentum), we have $k^2 \gg \left|
{{a}'' / a} \right|$ in Eq. (\ref{eq2}), i.e., term ${a}'' / a$ can
be neglected. In this case the general solution of Eq. (\ref{eq2})
has the usual oscillatory form [1,15]
\\

\begin{equation}
\label{eq3} \mu(\eta ) = A\exp ( - ik\eta ) + B\exp (ik\eta ).
\end{equation}
\\

Eqs. (\ref{eq1})-(\ref{eq3}) show that the high-frequency relic
GWs can be seen as the ``quasi-monochromatic waves'', their
amplitudes are the stochastic values containing the cosmology
scale factor $a(\eta )$. Of cause, for the EM response in
laboratory, we should use the intervals of laboratory time [i.e.,
$cdt = a(\eta )\eta $] and laboratory frequency [9]. Then Eq.
(\ref{eq1}) can be reduced to
\\

\begin{equation}
\label{eq4}
h(\vec {x},t) = A(k_g ) / a(t)\exp [i(\vec {k}_g \cdot \vec {x} - \omega _g
t)] + B(k_g ) / a(t)\exp [i(\vec {k}_g \cdot \vec {x} + \omega _g t)].
\end{equation}
\\

In the following we shall consider the EM resonant response
($\omega _e = \omega _g )$ in different cases. In figure 1 we draw
the symmetrical axis (the $z$-axis) of the Gaussian beam and the
propagating directions $\vec {k}_g $ of the arbitrary component of
the relic GWs.

$(a)\  \theta = 0$\textit{, i.e., the relic GW component propagates
along the positive direction of the z-axis.}

In this case, as we have shown [10] that the average value of the
$x$-component of the first-order perturbative photon flux (PPF) density is
\\
\begin{eqnarray}
n_x^{(1)} &=& - \frac{1}{{\hbar \omega _e }} \cdot \left\{
{\frac{{A_ \otimes  \hat B_y^{(0)} \psi _0 k_g y(z + l_1 )}}{{4\mu
_0 \left[ {1 + (z/f)^2 } \right]^{1/2} (z + f^2 /z)}}} \sin \left(
{\frac{{k_g r^2 }}{{2R}} - \tan ^{ - 1} \frac{z}{f}}
\right)\right.
\nonumber\\
 &+&  \frac{{A_ \otimes  \hat B_y^{(0)} \psi _0 y(z + l_1 )}}{{2\mu
_0 W_0^2 \left[ {1 + (z/f)^2 } \right]^{3/2} }} \left. {\cos
\left( {\frac{k_g r^2}{2R} - \tan ^{ - 1}\frac{z}{f}} \right)}
\right\} \cdot \exp \left( { - \frac{r^2}{W^2}} \right)
\nonumber\\
&-& \frac{1}{\hbar \omega _e }\left\{ {\left( {1 -
\frac{4x^2}{W^2}} \right)\frac{A_ \otimes \hat {B}_y^{(0)} \psi _0
k_g (z + l_1 )}{4\mu _0 R\left[ {1 + (z / f)^2} \right]^{1 / 2}}}
\right. \cdot \left[ {F_1 (y)\sin \left( {\frac{k_g x^2}{2R} -
\tan ^{ - 1}\frac{z}{f}} \right)} \right.
\nonumber\\
&+& \left. { F_2 (y)\cos \left( {\frac{{k_g x^2 }}{{2R}} - \tan ^{
- 1} \frac{z}{f}} \right)} \right] + \left[ {\frac{2}{{W^2 }} +
\left( {\frac{{k_g ^2 }}{{R^2 }} - \frac{1}{{W^4 }}} \right)x^2 }
\right]
\nonumber\\
&&\cdot \frac{{A_ \otimes  \hat B_y^{(0)} \psi _0 (z + l_1 )}}{{4\mu
_0 \left[ {1 + (z/f)^2 } \right]^{1/2} }} \left[ {F_1 (y)\cos \left(
{\frac{{k_g x^2 }}{{2R}} - \tan ^{ - 1} \frac{z}{f}} \right)}\right.
\nonumber\\
&-& F_2 (y) \left. \left. { {\sin \left( {\frac{k_g x^2}{2R} -
\tan ^{ - 1}\frac{z}{f}} \right)}} \right] \right\} \exp \left( {
- \frac{x^2}{W^2}} \right), (l_1 \le z \le l_2 )
\end{eqnarray}

\noindent
where
\\

\[
F_1 (y) = \int {\exp ( - \frac{y^2}{W^2})\cos (\frac{k_g y^2}{2R})dy,}
\]

\begin{equation}
\label{eq5}
F_2 (y) = \int {\exp ( - \frac{y^2}{W^2})\sin (\frac{k_g y^2}{2R})dy,}
\end{equation}
\\

\noindent are the quasi-probability integrals,$A_ \otimes = A /
a(t)$ is the stochastic value of the amplitude of the relic GW in
the laboratory frame of reference, $\hat {B}_y^{(0)} $ is the
background static magnetic field, which is localized in the region $
- l_1 \le z \le l_2 $, $\psi _0 $ is the amplitude of electric field
of the Gaussian beam, $W_0 $ is its minimum spot radius, $f = \pi
W_0^2 / \lambda _e $, $W = W_0 \left[ {1 + (z / f)^2} \right]^{1 /
2}$, and $R = z + f^2 / z$ is the curvature radius of the wavefronts
of the Gaussian beam at $z$ (see. e.g., Ref. [16]). We will show
that the PPF expressed by Eq. (5) has best perturbative effect than
that generated by the relic GW components propagating along other
directions.

$(b)\  \theta = \pi $\textit{, i.e., the relic GW component
propagates along the negative direction of the z-axis}.

By using the similar means, one finds
\\

\begin{eqnarray}
n_x^{(1)} &=& - \frac{1}{{\hbar \omega _e }} \cdot \left\{
{\frac{{A_ \otimes  \hat B_y^{(0)} \psi _0 k_g y(l_2-z )}}{{4\mu _0
\left[ {1 + (z/f)^2 } \right]^{1/2} (z + f^2 /z)}}} \sin \left(
{2kz+\frac{{k_g r^2 }}{{2R}} - \tan ^{ - 1} \frac{z}{f}}
\right)\right.
\nonumber\\
 &+&  \frac{{A_ \otimes  \hat B_y^{(0)} \psi _0 y(l_2-z )}}{{2\mu
_0 W_0^2 \left[ {1 + (z/f)^2 } \right]^{3/2} }} \left. {\cos \left(
{2kz+\frac{k_g r^2}{2R} - \tan ^{ - 1}\frac{z}{f}} \right)} \right\}
\cdot \exp \left( { - \frac{r^2}{W^2}} \right)
\nonumber\\
&-& \frac{1}{\hbar \omega _e }\left\{ {\left( {1 - \frac{4x^2}{W^2}}
\right)\frac{A_ \otimes \hat {B}_y^{(0)} \psi _0 k_g (l_2-z )}{4\mu
_0 R\left[ {1 + (z / f)^2} \right]^{1 / 2}}} \right. \cdot \left[
{F_1 (y)\sin \left( {2kz+\frac{k_g x^2}{2R} - \tan ^{ -
1}\frac{z}{f}} \right)} \right.
\nonumber\\
&+& \left. { F_2 (y)\cos \left( {2kz+\frac{{k_g x^2 }}{{2R}} - \tan
^{ - 1} \frac{z}{f}} \right)} \right] + \left[ {\frac{2}{{W^2 }} +
\left( {\frac{{k_g ^2 }}{{R^2 }} - \frac{1}{{W^4 }}} \right)x^2 }
\right]
\nonumber\\
&&\cdot \frac{{A_ \otimes  \hat B_y^{(0)} \psi _0 (l_2-z )}}{{4\mu
_0 \left[ {1 + (z/f)^2 } \right]^{1/2} }} \left[ {F_1 (y)\cos \left(
{2kz+\frac{{k_g x^2 }}{{2R}} - \tan ^{ - 1} \frac{z}{f}}
\right)}\right.
\nonumber\\
&-& F_2 (y) \left. \left. { {\sin \left( {2kz+\frac{k_g x^2}{2R} -
\tan ^{ - 1}\frac{z}{f}} \right)}} \right] \right\} \exp \left( { -
\frac{x^2}{W^2}} \right), (l_1 \le z \le l_2 )
\end{eqnarray}
\\

Different from Eq. (5), each and all items in Eq. (7) contain
oscillating factor $2kz$. We emphasize that $2kz \approx 419z$ for
the high-frequency relic GW of $\nu _g = 10^{10}\texttt{Hz}$,
namely, the factor $2kz$ will play major role in the region of the
effective coherent resonance. In other words, the sign of
$n_x^{(\ref{eq1})} $ will be quickly oscillated and
quasi-periodically changed as the coordinate $z$ in the region. Thus
the total effective PPF passing through a certain ``typical
receiving surface'' will be much less than that generated by the
relic GW component propagating along the positive direction of the
$z$-axis, (see Eq. (5) and Table 1)

 $(c)\ \theta = \pi / 2,\mbox{ }\phi \mbox{ = 0}$ \textit{, i.e., the
propagating direction of the relic GW component is not only
perpendicular to the symmetrical axis of the Gaussian beam, but also
vertical to the static magnetic field }$\hat {B}_y^{(0)} .$

Here we assume that the dimension of the $x$-direction of $\hat {B}_y^{(0)} $
is localized in the region $ - l_3 \le x \le l_4 $. Utilizing the similar
means, the first-order perturbative EM fields generated by the direct
interaction of the relic GW with the static magnetic field can be given by
[11]
\\

\begin{eqnarray}
\tilde {E}_y^{(\ref{eq1})}& =& \frac{i}{2}A_ \oplus \hat {B}_y^{(0)}
k_g c(x + l_3 )\exp [i(k_g x - \omega _g t)] + \frac{1}{4}A_ \oplus
\hat {B}_y^{(0)} c\exp [i(k_g x + \omega _g t)],
\nonumber\\
\tilde {B}_z^{(\ref{eq1})} &=& \frac{i}{2}A_ \oplus \hat {B}_y^{(0)}
k_g (x + l_3 )\exp [i(k_g x - \omega _g t)] - \frac{1}{4}A_ \oplus
\hat {B}_y^{(0)} \exp [i(k_g x + \omega _g t)],
\nonumber\\
\tilde {E}_z^{(\ref{eq1})} &=& - \frac{1}{2}A_ \otimes \hat
{B}_y^{(0)} k_g c(x + l_3 )\exp [i(k_g x - \omega _g t)] +
\frac{i}{4}A_ \otimes \hat {B}_y^{(0)} c\exp [i(k_g x + \omega _g
t)],
\nonumber\\
\tilde {B}_y^{(\ref{eq1})} &=& \frac{1}{2}A_ \otimes \hat
{B}_y^{(0)} k_g (x + l_3 )\exp [i(k_g x - \omega _g t)] +
\frac{i}{4}A_ \otimes \hat {B}_y^{(0)} \exp [i(k_g x + \omega _g
t)], \nonumber\\(l_3 \le x \le l_4)
\end{eqnarray}
\\

In this case the coherent syncro-resonance ($\omega _e = \omega _g
)$ between the perturbative fields, Eq. (8), and the Gaussian beam
can be expressed as the following first-order PPF density, i.e.,
\\

\begin{equation}
\label{eq6}
n_x^{(\ref{eq1})} = \frac{1}{\mu _0 \hbar \omega _e }\left[ {\langle \tilde
{E}_y^{(\ref{eq1})} \tilde {B}_z^{(0)} \rangle + \langle \tilde {E}_y^{(0)} \tilde
{B}_z^{(\ref{eq1})} \rangle - \langle \tilde {E}_z^{(\ref{eq1})} \tilde {B}_y^{(0)} \rangle
} \right],
\end{equation}
\\

\noindent where $\tilde {B}_y^{(0)} $, $\tilde {B}_z^{(0)} $ are the
$y$- and $z$- components of the magnetic filed of the Gaussian beam,
respectively, the angular brackets denote the average over time.
Notice that we choose the Gaussian beam of the transverse electric
modes (TE), so $\tilde {E}_z^{(0)} = 0$. By using the same method,
we can calculate $n_x^{(\ref{eq1})} $, Eq. (\ref{eq6}). For example,
first term in Eq. (\ref{eq6}) can be written as
\\

\begin{eqnarray}
&&\frac{1}{\mu\hbar\omega_{e}}\langle\tilde{E}_y^{(1)}\tilde{B}_z^{(0)}\rangle
=\nonumber\\
&-&\frac{1}{\hbar\omega_{e}}\left\{ {\frac{{A_ \oplus \hat B_y^{(0)}
\psi _0 k_g y(z + l_3 )}}{{4\mu _0 \left[ {1 + (z/f)^2 }
\right]^{1/2} (z + f^2 /z)}}} \sin \left( {k_g(x-z)+\frac{{k_g r^2
}}{{2R}} - \tan ^{ - 1} \frac{z}{f}} \right)\right.
\nonumber\\
&+&  \frac{{A_ \oplus \hat B_y^{(0)} \psi _0 y(z + l_3 )}}{{2\mu _0
W_0^2 \left[ {1 + (z/f)^2 } \right]^{3/2} }} \left. {\cos \left(
{k_g(x-z)+\frac{k_g r^2}{2R} - \tan ^{ - 1}\frac{z}{f}} \right)}
\right\}  \exp \left( { - \frac{r^2}{W^2}} \right)
\nonumber\\
&-& \frac{1}{\hbar \omega _e }\left\{ {\left( {1 - \frac{4x^2}{W^2}}
\right)\frac{A_ \oplus \hat {B}_y^{(0)} \psi _0 k_g y(z + l_3
)}{4\mu _0 R\left[ {1 + (z / f)^2} \right]^{1 / 2}}} \right. \cdot
\left[ {F_1 (y)\sin \left( {k_g(x-z)+\frac{k_g x^2}{2R} - \tan ^{ -
1}\frac{z}{f}} \right)} \right.
\nonumber\\
&+& \left. { F_2 (y)\cos \left( {k_g(x-z)+\frac{{k_g x^2 }}{{2R}} -
\tan ^{ - 1} \frac{z}{f}} \right)} \right] + \left[ {\frac{2}{{W^2
}} + \left( {\frac{{k_g ^2 }}{{R^2 }} - \frac{1}{{W^4 }}} \right)x^2
} \right]
\nonumber\\
&& \frac{{A_ \oplus  \hat B_y^{(0)} \psi _0 y(z + l_3 )}}{{4\mu _0
\left[ {1 + (z/f)^2 } \right]^{1/2} }} \left[ {F_1 (y)\cos \left(
{k_g(x-z)+\frac{{k_g x^2 }}{{2R}} - \tan ^{ - 1} \frac{z}{f}}
\right)}\right.
\nonumber\\
&-& F_2 (y) \left. \left. { {\sin \left( {k_g(x-z)+\frac{k_g
x^2}{2R} - \tan ^{ - 1}\frac{z}{f}} \right)}} \right] \right\} \exp
\left( { - \frac{x^2}{W^2}} \right), (l_3 \le z \le l_4 ).
\end{eqnarray}
\\

It can be shown that calculation for the 2nd and 3rd terms in Eq. (\ref{eq6}) is
quite similar to first term, and they have the same orders of magnitude, we
shall not repeat it here. Eq. (8) shows that the $n_x^{(\ref{eq1})} $ have a space
accumulation effect ($n_x^{(\ref{eq1})} \propto x)$. This is because the GWs
(gravitons) and EM waves (photons) have the same propagating velocity, so
that the two waves can generate an optimum coherent effect in the
propagating direction. However, unlike $n_x^{(\ref{eq1})} $ produced by the relic GW
component propagating along the positive direction of the $z$-axis [see, Eq.
(5)], the phase functions in Eq. (10) contain oscillating factor $k_g (x -
z)$, and it is always possible to choose $l_1 + l_2 \gg l_3 + l_4 $, i.e.,
the dimension of the $z$-direction of $\hat {B}_y^{(0)} $ is much larger than
its $x$-direction dimension. Because of such reasons, the PPF expressed by Eq.
(10) will be much less than that repressed by Eq. (5) (see, Table 1).

$(d)\  \theta = \pi / 2,\mbox{ }\phi = \pi / 2$\textit{, i.e., the
relic GW component propagates along the y-axis, which is parallel
with the static magnetic field }$\hat {B}_y^{(0)} .$

According to the Einstein-Maxwell equations of the weak fields, then the
perturbation of the GW to the static magnetic field vanishes [17], i.e.,
\\

\begin{equation}
\label{eq7}
n_x^{(\ref{eq1})} = 0.
\end{equation}
\\

It is very interesting to compare $n_x^{(\ref{eq1})} $ in Eqs. (5),
(7), (10) and (\ref{eq7}). As is shown that although they all
represent the PPFs propagating along the $x$-axis, their physical
behaviors are quite different. In the case of $\theta = \phi = \pi /
2$, $n_x^{(\ref{eq1})} = 0$, Eq. (\ref{eq7}); when $\theta = \pi $
and $\theta = \pi / 2,\mbox{ }\phi = 0$, the PPFs contain the
oscillating factors $2k_g z$ and $k_g (x - z)$, respectively [see
Eqs. (7) and (10)]. Only under the condition $\theta = 0$, the PPF,
Eq. (5), does not contain any oscillating factor, but only slow
enough variation function in the \textit{z} direction. This means
that $n_x^{(\ref{eq1})} $ produced by the relic GW component
propagating along the positive direction of the \textit{z}-axis, has
the best space accumulation effect. Thus our EM system would be very
sensitive to the propagating directions of the relic GWs. In other
words the EM system has a strong selection capability to the
resonant components from the stochastic relic GW background.

The total PPF passing through a certain ``typical receiving surface'' at the
\textit{yz}-plan will be
\\

\begin{equation}
\label{eq8} N_x^{(\ref{eq1})} = \int\!\!\!\int\limits_{\triangle S}
{\left. {n_x^{(\ref{eq1})} } \right|_{x = 0} } dydz.
\end{equation}
\\

In order to compare the PPFs generated by the different components of the
relic GWs, we introduce the typical laboratorial and cosmological
parameters:

(\ref{eq1}). $\Omega _{gw} \sim 5\times 10^{ - 6}$, the peak value
of the normalized energy density of the high-frequency relic GW
($\nu _g = 10^{10}\texttt{Hz})$ in the QIM [1]. Then $h\sim
\frac{\nu _H }{\nu }(\Omega _{gw} )^{1 / 2}\sim 4.48\times 10^{ -
31}$ [7], i.e., $A_ \oplus ,\mbox{ }A_ \otimes \sim 4.48\times 10^{
- 31}$, where $\nu _H \approx 2\times 10^{ - 18}\texttt{Hz}$ is the
Hubble frequency.

(\ref{eq2}). $P = 10\texttt{W}$, The power of the Gaussian beam. In
this case $\psi _0 \approx 1.26\times 10^3\texttt{Vm}^{ - 1}$ for
the Gaussian beam of $W_0 = 0.05\texttt{m}$.

(\ref{eq3}). $\hat {B}_y^{(0)} = 3\texttt{T}$, the strength of the
background static magnetic field.

(\ref{eq4}). $0 \le y \le W_0 $, $0 \le z \le 0.3\texttt{m}$, the
integration region $\bigtriangleup S$ (the receiving surface of the
PPF) in Eq. (\ref{eq8}), i.e., $\bigtriangleup S \approx 10^{ -
2}\texttt{m}^2$.

(5). $z = l_2 + l_1 = 4\texttt{m}$ and $x = l_3 + l_4 =
0.2\texttt{m}$, (i.e., $\bigtriangleup z \gg \bigtriangleup x)$ they
are the interacting dimensions between the relic GWs and the static
magnetic field in the $z$ and $x $ directions, respectively.

From the above parameters and Eqs. (5), (7), (10) and (\ref{eq8}), we obtain the
values of the PPFs as listed in Table 1.

Table 1 shows that the PPF produced by the relic GW component propagating
along the positive direction of the symmetrical axis of the Gaussian beam,
has a best resonant effect, i.e., largest perturbation and a good space
accumulation effect.

As for the distinction between the PPFs and the background photon fluxes, as
we have shown [10-12] that utilizing their very different physical behavior
in some local regions, they can be split by the special fractal membranes
[13,14], so that, the PPFs, in principle, would be observable.

Finally, it should be pointed out that superposition of the relic GWs
stochastic components will cause the fluctuation of the PPFs, even if such
``monochromatic components'' all satisfy the frequency resonant condition
($\omega _g = \omega _e )$. However, Eqs. (5), (7), (10) and (\ref{eq8}) show that
the metric fluctuation only causes the change of the instantaneous values of
the PPFs and does not influence the ``direction resonance''. i.e., it does
not influence the selection capability of the EM system to the propagating
directions of the relic GWs, and it does not influence average effect over
time of the PPFs.

Therefore, the high-frequency relic GWs satisfying the above
parameters requirement ($h\sim 10^{ - 31}$ or larger), frequency
resonance ($\nu_e = \nu_g = 10^{10}\texttt{Hz})$ and ``direction
resonance'', in principle, would be selectable and measurable in
seconds. More detailed schemes will be studied elsewhere.

\begin{figure}[htbp]
\centerline{\includegraphics[width=5cm]{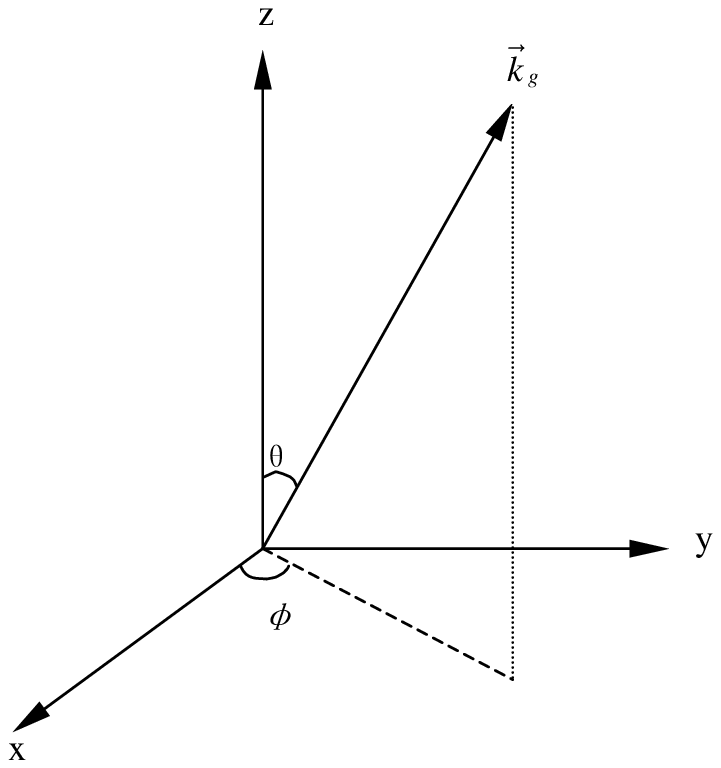}} \caption{The
\textit{z}-axis is the symmetrical axis of the Gaussian beam,
$\vec{k}_{g}$ represents the instantaneous propagating direction of
the arbitrary component of the relic GW.} \label{fig1}
\end{figure}

\begin{table}[htbp]
\caption {The PPFs generated by the resonant relic GW components
propagating along the different directions, here $\hat {B}^{(0)} =
3\texttt{T}$, $A_ \otimes ,\mbox{ }A_ \oplus \sim 4.48\times 10^{ -
31}$, $\nu_{g} = 10^{10}$Hz}

\center\begin{tabular}{c c} \hline The propagating directions&
The PPFs (s$^{ - 1})$ \\
\hline\\ $z$&
4.42$\times $10$^{2}$ \\
\\
$-z$&
1.07 \\
\\
$x$&
1.42$\times $10$^{-1}$ \\
\\
$y$&
0 \\
\hline
\end{tabular}
\label{tab1}
\end{table}

\end{document}